\documentclass[useAMS,usenatbib]{mn2e}
\usepackage{epsfig}
\usepackage{amsmath}

\newcommand{\be}{\begin{equation}}
\newcommand{\beq}{\begin{equation}}
\newcommand{\ba}{\begin{eqnarray}}
\newcommand{\ee}{\end{equation}}
\newcommand{\eeq}{\end{equation}}
\newcommand{\ea}{\end{eqnarray}}

\newcommand{\apj}{ApJ}
\newcommand{\apjl}{ApJL}
\newcommand{\mnras}{MNRAS}

\def\lsim{~\rlap{$<$}{\lower 1.0ex\hbox{$\sim$}}}

\def\gsim{~\rlap{$>$}{\lower 1.0ex\hbox{$\sim$}}}

\title[Correlation between high redshift galaxies and redshifted 21cm
emission]{The correlation between the distribution of galaxies and 21cm
emission at high redshifts}

\author[Wyithe \& Loeb]{J. Stuart B. Wyithe$^1$ and Abraham Loeb$^2$\\$^1$
School of Physics, University of Melbourne, Parkville, Victoria,
Australia\\$^2$ Harvard-Smithsonian Center for Astrophysics, 60 Garden St.,
Cambridge, MA 02138\\Email: swyithe@physics.unimelb.edu.au,
loeb@cfa.harvard.edu}

\begin{document}

%\date{\today}
%\pagerange{\pageref{firstpage}--\pageref{lastpage}} \pubyear{2006}

\maketitle

\label{firstpage}
\begin{abstract}

Deep surveys have recently discovered galaxies at the tail end of the
epoch of reionization. 
In the near future, these discoveries will be complemented by a new generation of low-frequency radio observatories that
will map the distribution of neutral hydrogen in the intergalactic medium through
its redshifted 21cm emission.  In this paper we calculate the expected cross-correlation
between the distribution of galaxies and intergalactic 21cm emission at
high redshifts.  We demonstrate using a simple model that overdense regions
are expected to be ionized early as a result of their biased galaxy formation. 
This early phase leads to an anti-correlation between the 21cm
emission and the overdensities in galaxies, matter, and neutral
hydrogen. Existing Ly$\alpha$ surveys probe galaxies that are highly
clustered in overdense regions. By comparing 21cm emission from regions
near observed galaxies to those away from observed galaxies, future
observations will be able to test this generic prediction and calibrate the
ionizing luminosity of high-redshift galaxies.
  
\end{abstract}


\section{Introduction}

An important question that should be addressed by successful models of
reionization concerns whether overdense or underdense regions become
ionized first.  In regions that are overdense, galaxies will be
over-abundant for two reasons; first because there is more material per
unit volume to make galaxies, and second because small-scale fluctuations
need to be of lower amplitude to form a galaxy when embedded in a
larger-scale overdensity (the so-called {\it galaxy bias}; see Mo \& White~1996). Regarding reionization of the intergalactic medium (IGM), the
first effect will be compensated by the increased density of gas to be
ionized. Furthermore, the increase in the recombination rate in overdense
regions will be counteracted by the galaxy bias in overdense regions. However the
latter effects need not cancel, and could either lead to enhanced or
delayed reionization in overdense regions.

The process of reionization also contains several layers of feedback.
Radiative feedback heats the IGM and results in the suppression of low-mass
galaxy formation (Efstathiou, 1992; Thoul \& Weinberg~1996; Quinn et al.~1996; Dijkstra et al.~2004). This delays the completion of reionization by
lowering the local star formation rate, but the effect is counteracted in overdense regions by
the biased formation of massive galaxies. The
radiation feedback may therefore be more important in low-density regions
where small galaxies contribute more significantly to the ionizing flux.

\begin{figure*}
\includegraphics[width=15cm]{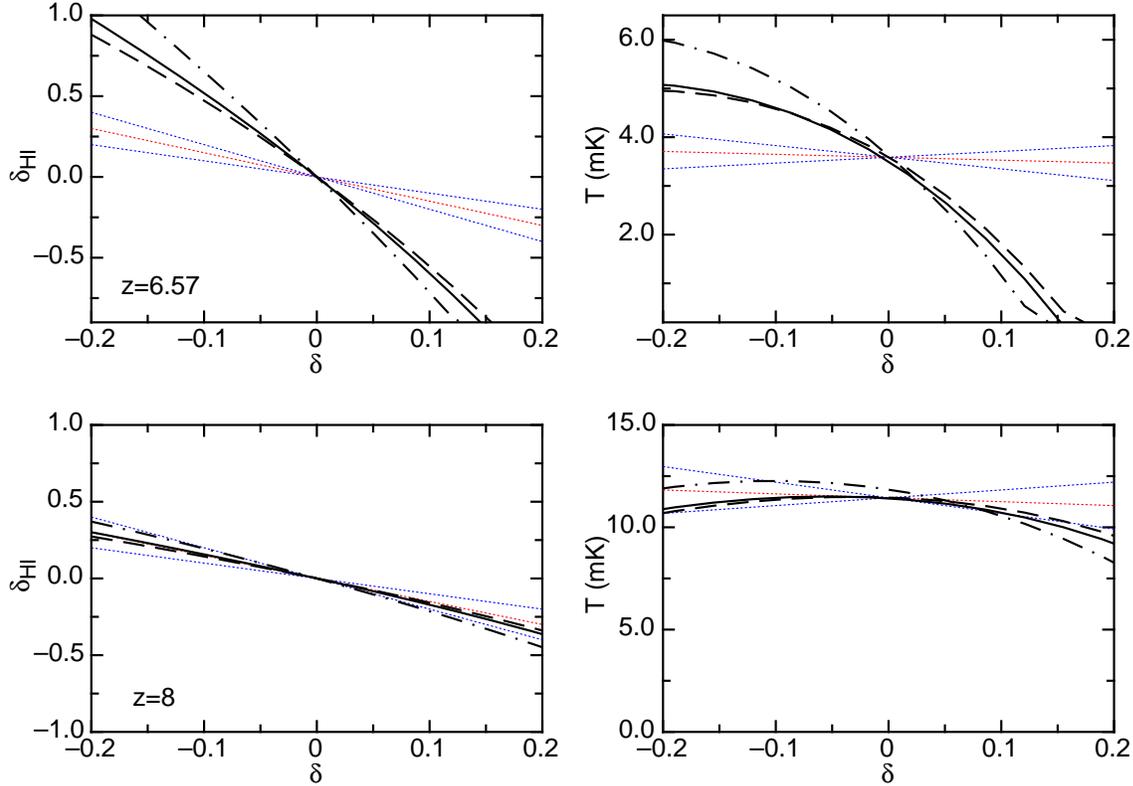} 
\caption{ \textit{Left}: Fluctuations in the hydrogen neutral fraction
[$\delta_{\rm HI}\equiv x_{\rm HI}(\delta)/\langle x_{\rm HI}\rangle-1$]
versus overdensity $\delta$ (for $R\gg R_{\rm min}$). The solid, dot-dashed
and dashed lines represent values of $C=10$, $C=2$ and $C=20$. For
comparison, we show three lines of slope $d\delta_{\rm HI}/d\delta=-0.5$,
$-1.5$ and $-2.5$. \textit{Right}: The dependence of the predicted
brightness temperature ($T$) on overdensity. The solid, dot-dashed and
dashed lines represent values of $C=10$, $C=2$ and $C=20$. For comparison,
the three lines show Equation~(\ref{neut}) with values of $\beta=-0.5$,
-1.5 and -2.5. The upper and lower rows correspond to observations at
$z=6.57$ and $z=8$ respectively. }
\label{fig1}
\end{figure*}

In this paper we use a simple model to evaluate the relative significance
of the above effects and compute the correlation between the local
overdensity and quantities such as the hydrogen neutral-fraction, the
brightness temperature of redshifted 21cm emission, and the overdensity of
massive galaxies.  In particular, we evaluate the prospects for finding
whether reionization is enhanced or suppressed in overdense regions based
on the combination of high redshift galaxy surveys and redshifted 21cm
surveys.

For illustration, the problem may be parameterised by relating the
neutral fraction [$x_{\rm HI}^{R}(\delta)$] in regions of linear
overdensity $\delta$ (smoothed over spheres of large radius $R$), to 
the average neutral fraction ($x_{\rm HI}$) using a power-law with index
$\beta(R)$
\begin{equation}
\label{neut}
x_{\rm HI}^{R}(\delta)=x_{HI}\left(1+\delta\right)^{\beta(R)}\approx
x_{HI}\left[1+\beta(R)\delta\right],
\end{equation} 
where we assume $\delta\ll 1$.
The resulting deviation in the brightness temperature of a region of
overdensity $\delta$ at $z\ga 6$ is
\begin{eqnarray}
\label{temp}
\nonumber
T(\delta)&\approx&22\mbox{mK}x_{\rm HI}^{R}\left(1+\frac{4}{3}\delta\right)\\
&\approx&22\mbox{mK}x_{HI}\left[1+\left(\frac{4}{3}+\beta(R)\right)\delta\right],
\end{eqnarray}
where we have assumed hydrogen in the IGM to have a spin temperature well in excess of the CMB temperature.
In Equation~(\ref{temp}) the pre-factor of 4/3 on the overdensity refers to
the spherically averaged enhancement of the brightness temperature due to peculiar velocities in overdense
regions (Bharadwaj \& Ali 2005; Barkana \& Loeb 2005). In a state where the
ionization fraction in the IGM is independent of density, the value of the
index is $\beta=0$ for all scales. If $\beta<0$, then ionization is
enhanced in overdense regions. Conversely, reionization is suppressed in
overdense regions if $\beta>0$.  High redshift galaxies are preferentially
located in large-scale regions with $\delta>0$. The departure of the mean
21cm brightness temperature in these regions from the average IGM provides the
opportunity to measure the value of $\beta$, and hence to determine whether
overdense or underdense regions were reionized first.  In \S 2 we describe
a simple model that predicts $\beta<0$. Regions of higher density
should therefore be more ionized and possess reduced levels of fluctuations
in redshifted 21cm emission.  We later describe how the dependence of 21cm
emission on overdensity could be extracted based on the fact that more
massive galaxies populate higher density regions. Throughout the paper we adopt the set of cosmological parameters determined
by {\it WMAP} (Spergel et al. 2006) for a flat $\Lambda$CDM universe.

\section{Simple model for the correlation of 21cm emission 
with large-scale overdensity}

\begin{figure*}
\includegraphics[width=15cm]{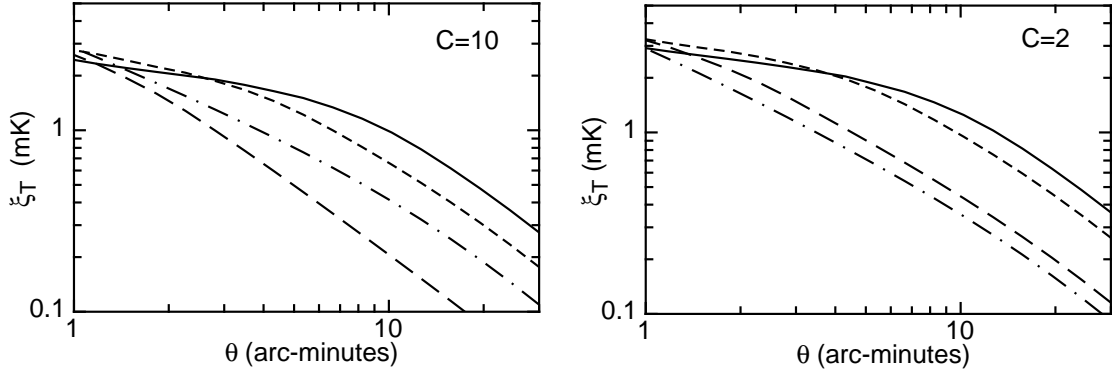}
\caption{ The auto-correlation function of 21cm brightness temperature
smoothed on different angular scales $\theta$. The four lines show the
results at redshifts $z=6.5$ (solid), $z=7$ (short-dashed), $z=8$ (long-dashed) and $z=10$
(dot-dashed). The left and right panels correspond to clumping factors $C=10$ and
$C=2$, respectively.}
\label{fig2}
\end{figure*} 

The evolution of the ionization fraction by mass $Q_{\delta,R}$ of a
particular region of scale $R$ with overdensity $\delta$ (at observed
redshift $z_{\rm obs}$) may be written as (Wyithe \& Loeb 2003)
\begin{eqnarray}
\label{history}
\nonumber
\frac{dQ_{\delta,R}}{dt} &=& \frac{N_{\rm ion}}{0.76}\left[Q_{\delta,R} \frac{dF_{\rm col}(\delta,R,z,M_{\rm ion})}{dt} \right.\\
\nonumber
&&\hspace{5mm}+ \left.\left(1-Q_{\delta,R}\right)\frac{dF_{\rm col}(\delta,R,z,M_{\rm min})}{dt}\right]\\
&-&\alpha_{\rm B}Cn_{\rm H}^0\left(1+\delta\frac{D(z)}{D(z_{\rm obs})}\right) \left(1+z\right)^3Q_{\delta,R},
\end{eqnarray}
where $N_{\rm ion}$ is the number of photons entering the IGM per baryon in
galaxies, $\alpha_{\rm B}$ is the case-B recombination co-efficient, $C$ is
the clumping factor (which we assume, for simplicity, to be constant), and
$D(z)$ is the growth factor between redshift $z$ and the present time. The
production rate of ionizing photons in neutral regions is assumed to be
proportional to the collapsed fraction $F_{\rm col}$ of mass in halos above
the minimum threshold mass for star formation ($M_{\rm min}$), while in
ionized regions the minimum halo mass is limited by the Jeans mass in an
ionized IGM ($M_{\rm ion}$). We assume $M_{\rm min}$ to correspond to a
virial temperature of $10^4$K, representing the hydrogen cooling threshold,
and $M_{\rm ion}$ to correspond to a virial temperature of $10^5$K,
representing the mass below which infall is suppressed from an ionized IGM
(Dijkstra et al.~2004). In a region of co-moving radius $R$ and mean
overdensity $\delta(z)=\delta D(z)/D(z_{\rm obs})$ [specified at redshift $z$ instead of the usual
$z=0$], the relevant collapsed fraction is obtained from the extended
Press-Schechter~(1974) model (Bond et al.~1991) as
\begin{equation}
F_{\rm col}(\delta,R,z) = \mbox{erfc}{\left(\frac{\delta_{\rm
c}-\delta(z)}{\sqrt{2\left(\left[\sigma_{\rm
gal}\right]^2-\left[\sigma(R)\right]^2\right)}}\right)},
\end{equation}
where $\mbox{erfc}(x)$ is the error function, $\sigma(R)$ is the variance
of the density field smoothed on a scale
$R$, and $\sigma_{\rm gal}$ is the variance of the density field smoothed
on a scale $R_{\rm gal}$, corresponding to a mass scale of $M_{\rm min}$ or $M_{\rm ion}$ (both evaluated at redshift $z$ rather than at $z=0$).  In
this expression, the critical linear overdensity for the collapse of a
spherical top-hat density perturbation is $\delta_c\approx
1.69$. The details of the galaxy bias and radiative feedback must
be incorporated in order to compute how $F_{\rm col}$ and $C$ vary with
$\delta$. Before proceeding, we stress that this model is intended to be
illustrative only. While semi-analytic models can offer a picture of the
global properties of reionization, the detailed processes must be modeled
numerically using cosmological simulations (e.g. Zahn et al.~2006).

\begin{figure*}
\includegraphics[width=15cm]{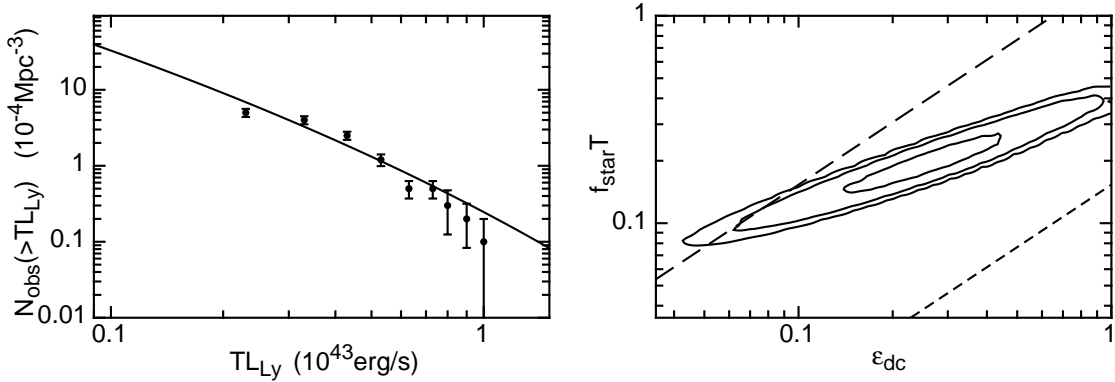} \caption{\textit{Left panel:} The
 number counts of Ly$\alpha$ emitters as a function of observed luminosity $\mathcal{T}L_{\rm Ly}$. The data points are from the
 cumulative number counts of Kashikawa et al.~(2006) for the SDF
 (incompleteness corrected). The solid line is the best fit model assuming
 Equations~(\ref{Lyalpha}-\ref{density}), with $\epsilon_{\rm dc}= 0.38$
 and $f_{\rm star}\mathcal{T}=0.24$. \textit{Right panel:} contours of likelihood in
 the $f_{\rm star}\mathcal{T}$ versus $\epsilon_{\rm dc}$ plane at 14\%,
 26\% and 64\% of the maximum likelihood. These contours correspond to the
 3, 2 and 1-sigma levels of a Gaussian distribution. The long and short
 dashed lines show lines of constant mass at $M=10^{10}M_\odot$ and
 $M=10^{11}M_\odot$ respectively.}
\label{fig3}
\end{figure*} 

Equation~(\ref{history}) may be integrated as a function of $\delta$ at the
observed redshift. As an example, we find the value of $N_{\rm ion}$ that
yields overlap of ionized regions at the mean density IGM by $z\sim6$
(White et al.~2003). We then use the model to compute the filling fraction
of ionized regions at $z=6.57$ (corresponding to the redshift of Ly$\alpha$
galaxy surveys discussed in the next section) and at a higher redshift of
$z=8$. The results are shown in the left panels of Figure~\ref{fig1}, where
fluctuations in neutral fraction [$\delta_{\rm HI}\equiv x_{\rm
HI}(\delta)/\langle x_{\rm HI}\rangle-1$] are plotted versus
$\delta$. Results are shown assuming clumping factors of $C=10$ (solid
lines), $C=2$ (dot-dashed lines), and $C=20$ (dashed lines). Here the
overdensities correspond to length scales in the IGM that are significantly
in excess of the length scale corresponding to the minimum mass
(i.e. Figure~\ref{fig1} represents scales where $[\sigma_{\rm
min}]^2-[\sigma(R)]^2\approx \sigma_{\rm min}^2)$. Therefore while
Figure~\ref{fig1} shows dependencies out to values of $\delta\sim0.5$, such
overdensities are rare on these large scales. We find that overdense
regions are more highly ionized than underdense regions, implying that the
bias of massive galaxies in overdense regions dominates over the increased
recombination rate there. Figure~\ref{fig1} shows that fluctuations in the
neutral fraction at a fixed overdensity (as measured at an observed
redshift $z_{\rm obs}$) are smaller at earlier times. The reason is simply
that the neutral fraction is larger at earlier times, so that fluctuations
in ionization fraction result in smaller relative fluctuations in neutral
fraction. The left panels of Figure~\ref{fig1} also show three lines of
slope $d\delta_{\rm HI}/d\delta=-0.5$, $-1.5$ and $-2.5$. From comparison
of these lines with our model prediction, we find that the value of $\beta$
in Equations~(\ref{neut}) and (\ref{temp}) is around $\beta\sim-1.5$ at
$z=8$ (for $C=10$), and is significantly steeper at lower redshifts or
lower clumping factors. Figure~\ref{fig1} suggests that a sufficiently large
clumping factor would result in an ionization fraction that was lower in
overdense relative to underdense regions.

On the right hand panels of Figure~\ref{fig1} we plot the variation of
predicted 21cm brightness temperature ($T$) with overdensity. Results are
again shown assuming clumping factors of $C=10$ (solid lines), $C=2$
(dot-dashed lines) and $C=20$ (dashed lines). For comparison, the three lines show
Equation~(\ref{temp}) with values of $\beta=-0.5$, -1.5 and -2.5. At
positive overdensities, we find a reduced neutral fraction. However in
these regions, the neutral hydrogen present is at an increased
overdensity. The brightness temperature of an overdense region is therefore
reduced relative to the average by increased ionization, but increased
relative to average by the higher gas density.  Conversely the brightness
temperature of an underdense region is increased relative to the average by
reduced ionization, but decreased relative to average by the lower gas
density. At large redshifts and/or clumping factor, we find that the sum of these effects results in a non-monotonic
dependence of $T$ on $\delta$, with brightness temperature peaking near
$\delta=0$.

On co-moving scales ($R$) sufficiently large that the variance in the
density field remains in the linear regime, we are able to compute the
auto-correlation function [$\xi_T(\theta)$] of brightness temperature
smoothed with top-hat windows of angular radius $\theta=R/d_{\rm A}(z)$,
where $d_{\rm A}$ is the angular diameter distance:
\begin{eqnarray}
\nonumber
\xi_T(\theta) &=& \langle \left(T-\langle T\rangle\right)^{2}\rangle^{1/2}\\
 &=& \left[\frac{1}{\sqrt{2\pi}\sigma(R)}\int d\delta \left(T(\delta)-\langle T\rangle\right)^2e^{-\frac{\delta^2}{2\sigma(R)^2}} \right]^{\frac{1}{2}},
\end{eqnarray}
where 
\begin{equation}
\langle T\rangle = \frac{1}{\sqrt{2\pi}\sigma(R)}\int d\delta~ T(\delta)e^{-\frac{\delta^2}{2\sigma(R)^2}} ,
\end{equation}
and $\sigma(R)$ is the variance of the density field (at redshift $z$) smoothed on a scale
$R$. The resulting auto-correlation functions are shown in
Figure~\ref{fig2} assuming clumping factors of $C=10$ (left) and $C=2$
(right). The four curves correspond to redshifts of $z=6.5$, 7, 8 and 10,
and demonstrate (in the case of $C=10$) that the amplitude of the auto-correlation function need
not vary monotonically with redshift (or average neutral fraction).

\section{The masses of Ly$\alpha$ emitters}

\label{Lymass}

We would like to determine whether the distribution of high redshift
galaxies can be combined with redshifted 21cm maps to probe the correlations
between the redshifted 21cm emission and galaxies, and hence between the ionization fraction and the large-scale overdensity. We
concentrate on the case of Ly$\alpha$ emitting galaxies. In order to
estimate the bias of Ly$\alpha$ emitting galaxies relative to the
underlying density field, we first compare the observed abundance of
Ly$\alpha$ emitters to a simple model in order to estimate their galaxy
mass.

In the Subaru Deep Field (SDF) Kashikawa et al.~(2006) find the density
[$N_{\rm obs}(>\mathcal{T}L_{\rm Ly})$] of Ly$\alpha$ emitters brighter than an
observed luminosity ($\mathcal{T}L_{\rm Ly}$) at $z=6.57$. Here $L_{\rm Ly}$ is the intrinsic luminosity of the emitter and $\mathcal{T}$ is the transmission of Ly$\alpha$ photons through the IGM\footnote{Kashikawa et al.~(2006) assume $\mathcal{T}=1$ when converting from observed flux to luminosity.}. In the lowest luminosity bin,
$\mathcal{T}L_{\rm Ly}=3\times10^{42}$erg/s, and the density is $N_{\rm obs}(>\mathcal{T}L_{\rm
Ly})=4\times10^{-4}$Mpc$^{-3}$. The number counts from Kashikawa et
al.~(2006) are reproduced in Figure~\ref{fig3} (for the case where the
densities are corrected for incompleteness). Our simple model for
Ly$\alpha$ emitters (Haiman \& Cen~2005) assumes the luminosity to be
\begin{equation}
\label{Lyalpha}
\mathcal{T}L_{\rm Ly}=3\times10^{42}\mbox{erg/s}\left(\frac{f_{\rm star}\mathcal{T}}{0.2}\right)\left(\frac{\epsilon_{\rm dc}}{0.1}\right)^{-1}\left(\frac{M}{10^{10}M_\odot}\right),
\end{equation}
where $\epsilon_{\rm dc}$ is the duty-cycle, $f_{\rm star}$ is the star-formation
efficiency, $M$ is the halo mass, and we have assumed the escape fraction
of ionizing photons to be much smaller than unity. The density of
Ly$\alpha$ emitters more luminous than $\mathcal{T}L_{\rm Ly}$ is given by
\begin{equation}
\label{density}
N(>\mathcal{T}L_{\rm Ly})=\epsilon_{\rm dc}\int_{M(L_{\rm Ly})}^\infty dM \frac{dn}{dM},
\end{equation}
where $dn/dM$ is the Press-Schechter~(1974) mass function (comoving number
density per galaxy mass) with the modification of Sheth \&
Tormen~(2002). For different combinations of $\epsilon_{\rm dc}$ and
$f_{\rm star}\mathcal{T}$ we then generate model number counts $N(>\mathcal{T}L_{\rm Ly})$ using
Equations~(\ref{Lyalpha}-\ref{density}). These number counts are compared
to the data from the SDF. For each parameter set, a likelihood is produced
$\mathcal{L}_{\rm Ly}(\epsilon_{\rm dc},f_{\rm
star}\mathcal{T})=\Pi_{i=1}^N\exp{\left[-\frac{1}{2}(N_{{\rm
obs},i}-N_i)^2/\sigma_i^2\right]}$, where $N_{{\rm obs},i}$ and $N_i$ are
evaluated in the $i$th observed luminosity bin ($i=1-9$), and $\sigma_i$ is
the uncertainty in observed density at the $i$th luminosity bin. In the
right panel of Figure~\ref{fig3} we show contours of the 14\%, 26\% and
64\% of the maximum likelihood, corresponding to the 3, 2 and 1-sigma
levels of a Gaussian distribution. The best fit model assuming
Equations~(\ref{Lyalpha}-\ref{density}) has $\epsilon_{\rm dc}= 0.38$ and
$f_{\rm star}\mathcal{T}=0.24$. We expect a transmission of order unity
(Dijkstra \& Wyithe~2006). These values are unexpectedly high, however the
model is degenerate over a wide range of parameter values. The best fit
model is plotted over the data in the left panel of Figure~\ref{fig3}. We
are interested in the mass of Ly$\alpha$ emitters. In the right hand panel
we show lines of constant mass corresponding to the lowest luminosity
($\mathcal{T}L_{\rm Ly}=3\times10^{42}$erg/s) bin. The long and short dashed lines
represent masses of $M=10^{10}M_\odot$ and $M=10^{11}M_\odot$
respectively. Based on these results we conclude that the halo masses of
Ly$\alpha$ emitters in the SDF are larger than $10^{10}M_\odot$. In the
remainder of this paper we show 21cm-galaxy correlations for both
$M=10^{10}M_\odot$ and $M=10^{11}M_\odot$.

\section{Cross-correlation of 21cm emission with galaxy properties}
\label{crosscorr}

\begin{figure*}
\includegraphics[width=15cm]{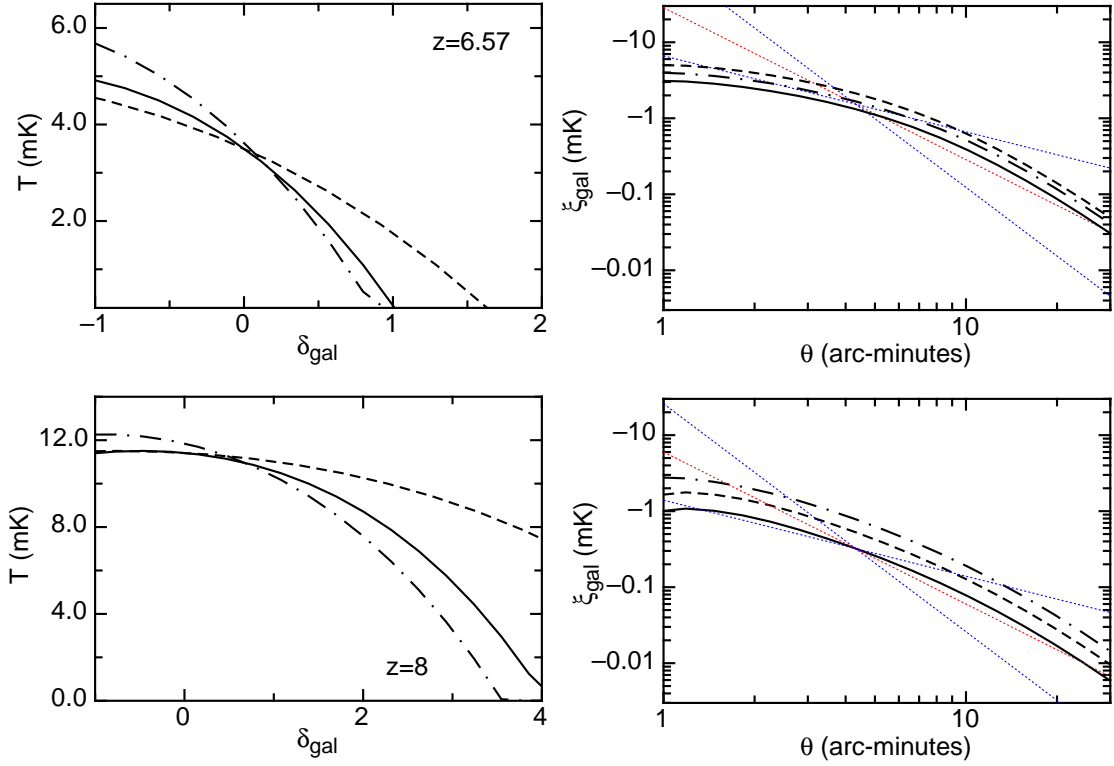} 
\caption{ \textit{Left}: 21cm brightness temperature as a function of
$\delta_{\rm gal}$. Two values of galaxy mass are assumed for a clumping of
$C=10$, $M=10^{10}M_\odot$ (solid line) and $M=10^{11}M_\odot$ (dashed
line). The dot-dashed line shows $C=2$ with
$M=10^{10}M_\odot$. \textit{Right}: The cross-correlation function
$\xi_{\rm gal}=\langle \delta_{\rm gal}\times(T-\langle T\rangle)\rangle$ for the
IGM smoothed on various angular scales ($\theta$). The function is
presented assuming $C=10$ for masses of $M=10^{10}M_\odot$ (solid line) and
$M=10^{11}M_\odot$ (dashed line). The dot-dashed line represents $C=2$
with $M=10^{10}M_\odot$. The lines show power-laws of slope
$d(\log\xi_{\rm gal})/d(\log\theta)=-1$, $-2$ and $-3$.  The upper and
lower rows correspond to observations at $z=6.57$ and $z=8$ respectively. }
\label{fig4}
\end{figure*} 

Next we discuss the cross-correlation between the number density of massive
galaxies with 21cm emission. The observed overdensity of galaxies is simply
$\delta_{\rm gal}=4/3\times b(M,z)\delta$, where $b(M, z)$ is the galaxy
bias, and the pre-factor of 4/3 arises from a spherical average over the
infall peculiar velocities (Kaiser 1987).  The value of bias $b$ for a halo
mass $M$ may be approximated using the Press-Schechter formalism (Mo \&
White~1996), modified to include non-spherical collapse (Sheth, Mo \&
Tormen~2001)
\begin{eqnarray}
\label{bias}
\nonumber
b(M,z) = 1 &+& \frac{1}{\delta_{\rm c}}\left[\nu^{\prime2}+b\nu^{\prime2(1-c)}\right.\\
&&\hspace{10mm}-\left.\frac{\nu^{\prime2c}/\sqrt{a}}{\nu^{\prime2c}+b(1-c)(1-c/2)}\right],
\end{eqnarray}
where $\nu\equiv {\delta_{\rm c}^2}/{\sigma^2(M)}$,
$\nu^\prime\equiv\sqrt{a}\nu$, $a=0.707$, $b=0.5$ and $c=0.6$. Here $\sigma(M)$ is the variance of the density field smoothed on a mass scale $M$ at redshift $z$. This
expression yields an accurate approximation to the halo bias determined
from N-body simulations (Sheth, Mo \& Tormen~2001).

In the left-hand panels of Figure~\ref{fig4} we plot 21cm brightness
temperature as a function of $\delta_{\rm gal}$. Results are shown for two
values of galaxy mass, $M=10^{10}M_\odot$ (solid line) and
$M=10^{11}M_\odot$ (dashed line) assuming $C=10$. We also show results for
$M=10^{10}M_\odot$ assuming $C=2$ (dot-dashed line). The upper row of
Figure~\ref{fig4} shows results for $z=6.57$, and the lower row for
$z=8$.

The properties of the galaxy population correlate with the level of
redshifted 21cm emission. These properties depend on the overdensity of the IGM
whose typical fluctuation level is a function of scale. As a result the amplitude of
the correlation between galaxy overdensity and 21cm emission will therefore also
be dependent on angular scale. In the right panels of
Figure~\ref{fig4} we plot the cross-correlation function 
\begin{eqnarray}
\nonumber
\xi_{\rm
gal}(\theta)&=&\langle\delta_{\rm gal}\times(T-\langle T\rangle)\rangle\\
&=&\frac{1}{\sqrt{2\pi}\sigma(R)}\int d\delta\left(\delta_{\rm gal}\times(T-\langle T\rangle)\right)e^\frac{-\delta^2}{2\sigma(R)^2}
\end{eqnarray}
for the IGM
smoothed on various angular scales, $\theta=R/d_{\rm A}$. Results are again shown for
two values of galaxy mass, $M=10^{10}M_\odot$ (solid line) and
$M=10^{11}M_\odot$ (dashed line) assuming $C=10$. As before we also show
results for $M=10^{10}M_\odot$ assuming $C=2$ (dot-dashed line). The lines
show power-laws of slope $d(\log|\xi_{\rm gal}|)/d(\log\theta)=-1$, $-2$ and
$-3$ respectively. Since in these cases we find the brightness
temperature to be lower in over-dense regions, we also find that massive
galaxies correlate negatively with 21cm brightness temperature.
The variance is lower on larger scales, and so the amplitude of the
correlation is reduced.

The results shown in Figures~\ref{fig1}, \ref{fig2} and \ref{fig4} are
sensitive to the value of clumping factor. Weaker clumping
leads to greater ionization in overdense regions, and hence a larger
variation of ionization fraction with overdensity. This in turn leads to
increases in the amplitudes of the auto-correlation and cross-correlation
functions.

On scales of $\theta\sim3'$ the typical fluctuation in units of 1mK
multiplied by the typical overdensity of galaxies is of order unity. In the
next section we demonstrate that the brightness temperature around
Ly$\alpha$ emitters at $z\sim6$ should be sufficiently different from the
average value to be detectable by the first generation of redshifted 21cm surveys.

\begin{figure*}
\includegraphics[width=15cm]{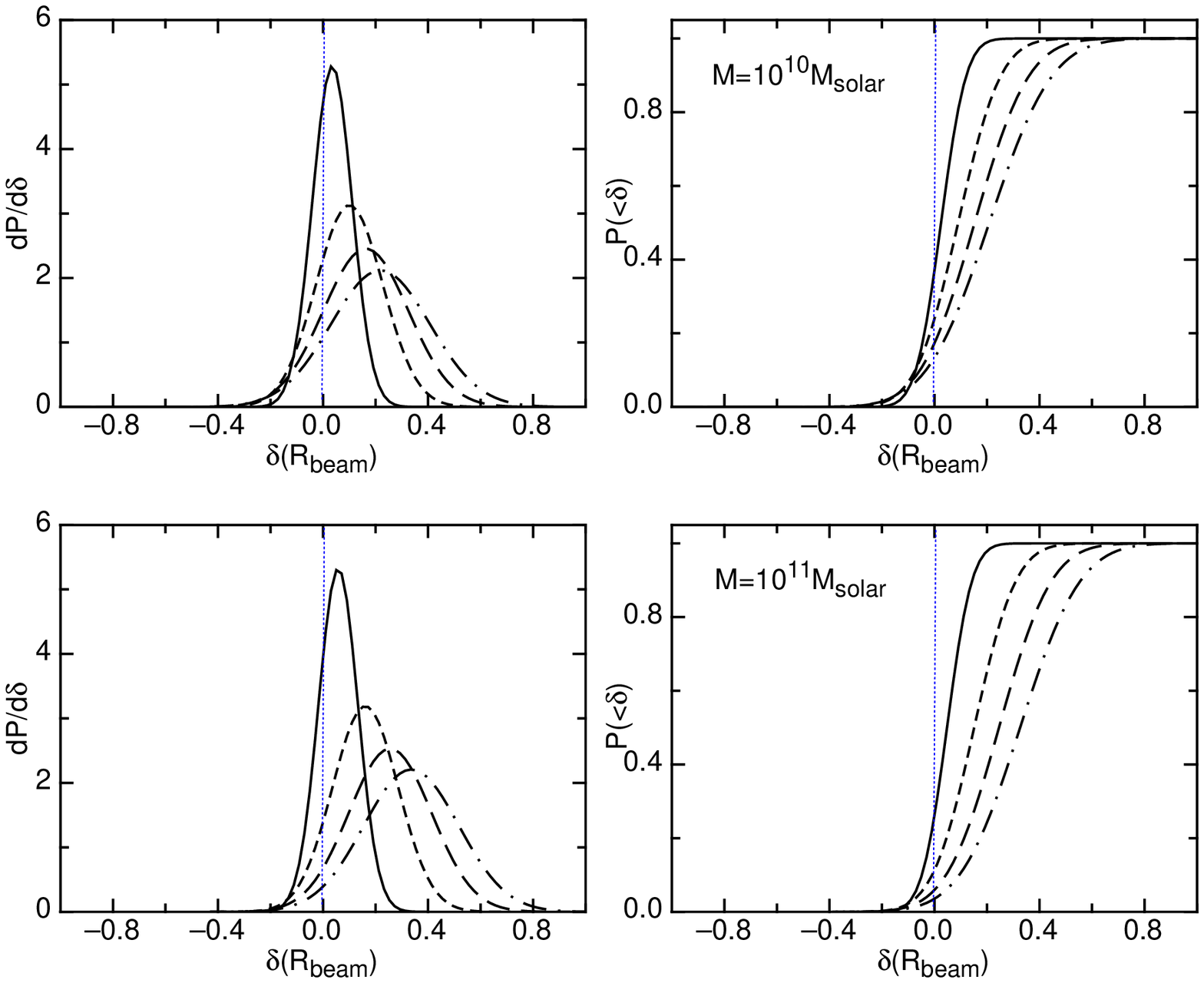} \caption{The differential (left) and
cumulative (right) probability distributions for the large scale overdensity at the
location of an observed Ly-alpha emitter. The solid, short-dashed, long-dashed and dot-dashed
lines correspond to splitting up the Ly$\alpha$ survey into $N_{\rm
bin}=1$, 2, 3 and 4 redshift bins. The upper and lower rows assume
Ly$\alpha$ galaxy masses of $10^{10}M_\odot$ and $10^{11}M_\odot$
respectively.}
\label{fig5}
\end{figure*}

\section{Measurement of correlations between Ly$\alpha$ emitters and Fluctuating 21cm emission}

The correlation between massive galaxies (which probe dense regions) and
the 21cm signal (which traces the geometry of the neutral gas) will offer a
direct probe of the process of reionization. The results of \S~\ref{Lymass}
and \S~\ref{crosscorr} suggest that Ly$\alpha$ emitters reside in massive
galaxies at high redshift, and that overdensities in the number counts of
these galaxies trace the more highly ionized regions. Observationally,
Ly$\alpha$ surveys may provide the most straightforward comparison with
21cm maps for several reasons which we outline below. As a concrete example
we consider the case of the Subaru Deep field and its comparison with an
MWA-LFD ({\it Mileura Wide Field Array-Low Frequency
Demonstrator}\footnote{see
http://www.haystack.mit.edu/ast/arrays/mwa/index.html}) field.  We can
estimate the sizes of fluctuations needed for detection by considering the
uncertainty in the 21cm brightness averaged over synthesised beams which do
or do-not contain Ly$\alpha$ emitting galaxies.

Radio telescopes measure images in data cubes and an instrument like the
MWA-LFD will have much higher resolution (spatially) along the line-of sight
than perpendicular to the line-of-sight. This allows for the option of
binning in frequency space so that the data cube can have equal resolution
in all three spatial dimensions. For an expected MWA-LFD resolution (beam radius) of $\theta_{\rm beam}=5$ arc-minutes, this
corresponds to a region of $\pm0.85$MHz or $\pm1.6$ physical Mpc along the line-of-sight at $z=6.5$. A
frequency interval of $2\times0.85$MHz$=1.7$MHz at $z=6.5$ corresponds to a value of $\Delta
z/(1+z)=0.008$, or a redshift range of $\Delta z=0.06$. For an
efficient cross-correlation we would therefore like galaxy redshifts to be known to $\pm0.03$, which requires spectroscopy. Surveys for Ly$\alpha$
emitters narrow down the redshift range initially via the use of narrow-band
filters, and then find sources with a strong emission line, making
spectroscopy attainable.

The characteristic lengths corresponding to the expected resolution of the
MWA-LFD are well matched to the dimensions of surveys for Ly$\alpha$
emitting galaxies. For example, the SDF has a total area of 876 square
arc-minutes. In this field the Subaru team found 50 Ly$\alpha$ emitter
candidates. Spectra of 22 of these were obtained of which 16 were verified
as high redshift galaxies, implying $N_{\rm Ly}\sim36$ galaxies in this deep
field. The line-of-sight dimension of the survey volume is set by the width
of the narrow-band filter ($\Delta\lambda=132$\AA), centered on a mean
wavelength of $\lambda_0=9196$\AA. This results in a field depth of $\Delta
z=(1+z)\Delta\lambda/\lambda_0=0.11$,
or $l=(cdt/dz)\Delta z\sim5.9$Mpc. At $z\sim6$ this depth is comparable to, but larger
than the line-of-sight dimension corresponding to the expected MWA-LFD angular
resolution. We may therefore divide the line of sight distance into a
number of bins $N_{\rm bins}$, so that the size of each bin is $l_{\rm
bin}=l/N_{\rm bin}$. We first define the angular scale that results in an
angular beam radius $\theta_{\rm beam}=R_{\rm beam}/d_{\rm A}$, where
$R_{\rm beam}=l_{\rm bin}/2$. This results in cylindrical volumes of space
where the diameter equals the length of the cylinder.  We are now able to
find the number of such cylindrical regions in the SDF survey volume
($N_{\rm total})$, as well as the number containing galaxies [$N_{\rm
gal}=N_{\rm Ly}/(1+\xi_{\rm Ly})$ where $\xi_{\rm Ly}$ is the excess
probability above random of finding a second galaxy within a cylinder], and
the number of cylinders not containing galaxies ($N_{\rm nogal}$).

We next discuss the response of a phased array to the brightness
temperature contrast of the IGM. Assuming that calibration can be performed
ideally, and that foreground subtraction is perfect, the root-mean-square
fluctuations in brightness temperature are given by the radiometer equation
\begin{equation}
\langle\Delta T^2\rangle^{1/2} = \frac{\epsilon\lambda^2T_{\rm sys}}{A_{\rm
tot}\Omega_{\rm b}\sqrt{t_{\rm int}\Delta\nu}},
\end{equation}
where $\lambda$ is the wavelength, $T_{\rm sys}$ is the system
temperature, $A_{\rm tot}$ the collecting area, $\Omega_{\rm b}$ the
effective solid angle of the synthesized beam in radians, $t_{\rm
int}$ is the integration time, $\Delta\nu$ is the size of the
frequency bin, and $\epsilon$ is a constant that describes the overall
efficiency of the telescope. We optimistically adopt $\epsilon=1$ in this
paper. In units relevant for upcoming telescopes and at $\nu=200$MHz,
we find (Wyithe, Loeb \& Barnes~2005)
\begin{eqnarray}
\label{noiseeqn}
\nonumber
\Delta T &=& 7.5\left(\frac{1.97}{C_{\rm beam}}\right)\mbox{mK} \left(\frac{A}{A_{\rm LFD}}\right)^{-1}\\
&\times&\left(\frac{\Delta\nu}{1\mbox{MHz}}\right)^{-1/2}\left(\frac{t_{\rm int}}{100\mbox{hr}}\right)^{-1/2}\left(\frac{\theta_{\rm beam}}{5^\prime}\right)^{-2}.
\end{eqnarray}
Here $A_{\rm LFD}$ is the collecting area of a phased array consisting of
500 tiles each with 16 cross-dipoles [the effective collecting area of an
LFD tile with $4\times4$ cross-dipole array with 1.07m spacing is
$\sim17-19$m$^2$ between 100 and 200MHz (B. Correy, private
communication)]. The system temperature at 200MHz will be dominated by the
sky and has a value $T_{\rm sys}\sim250$K.  $\Delta\nu$ is the frequency
range over which the signal is smoothed and $\theta_{\rm beam}$ is the size
of the synthesized beam. The value of $\theta_{\rm beam}$ can be regarded
as the radius of a hypothetical top-hat beam, or as the variance of a
hypothetical Gaussian beam. The corresponding values of the constant
$C_{\rm beam}$ are 1 and 1.97 respectively.  Given the noise per
synthesised beam, we can find the noise averaged in regions with and
without galaxies as $\Delta T_{\rm gal}=\Delta T/\sqrt{N_{\rm gal}}$ and
$\Delta T_{\rm nogal}=\Delta T/\sqrt{N_{\rm nogal}}$ respectively. The resulting
uncertainty in brightness temperature between regions with and without
galaxies is $\Delta T_{\rm diff}=\sqrt{(\Delta T_{\rm gal})^2+(\Delta
T_{\rm nogal})^2}$.

The measurement of fluctuations in brightness temperature between
over-dense and under-dense regions will require removal of emission from
unresolved foreground objects. At a fixed frequency and on scales of a few
arc-minutes, these foregrounds are believed to vary in amplitude at a level
several orders of magnitude in excess of the expected 21cm signal (Di~Matteo et al.~2002). However, the foregrounds are expected to have smooth
power-law spectra, while 21cm emission from the IGM will fluctuate in both
space and frequency. This smoothness should allow removal through
subtraction of a continuum component, leaving fluctuations due to 21cm
emission. However since the amplitude of the continuum will be different
along each line of sight, we will be unable to determine its absolute
level. Rather, the fluctuations in 21cm emission will need to be measured
relative to the average continuum component along each line of sight (about
which the fluctuations in brightness temperature will average to
zero). This average continuum must be measured from a region of spectrum
having finite length (the bandpass), and will therefore be determined only
to an accuracy corresponding to fluctuations in the mean 21cm emission
across the whole bandpass. For the MWA-LFD the bandpass will be 32MHz. We
may estimate the level of uncertainty introduced through
subtraction of the continuum by considering the variance of the density
field in cylinders of radius a few arc-minutes (the synthesised beam size),
and line of sight length corresponding to the band-pass. This variance can
be shown to be around a few percent, which should be compared to the
10-20\% representing the variance of the density field smoothed in spheres
of radius a few arc-minutes. Therefore, while fluctuations in the continuum
subtraction will add to the uncertainty in the measurement, they are
substantially smaller than the 21cm fluctuations of interest.

In summary there are two aspects of the problem. First, there is the question of the
distribution of overdensities (smoothed on the scale of the 21cm beam)
in the IGM surrounding high redshift galaxies. The mean of this distribution must be significantly
in excess of zero for any correlation of Ly$\alpha$ galaxies with large
scale fluctuations in 21cm emission to be detectable. Second, there is the
question of the sensitivity of the radio array, and the difference
in average brightness temperature between regions (on the scale of the
beam) that occupy or do not occupy galaxies. We treat each of these issues
in turn.

\begin{figure*}
\includegraphics[width=15cm]{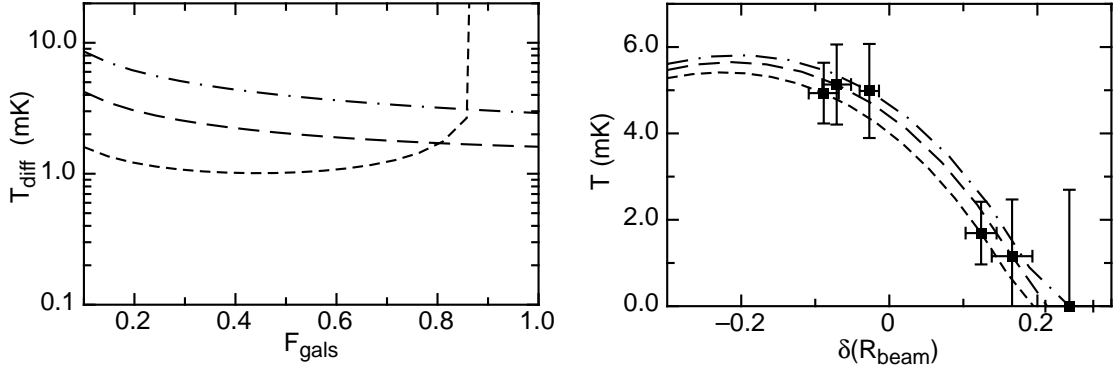} \caption{\textit{Left panel:} The
measured error in brightness temperature between regions that contain or do not contain
Ly$\alpha$ emitters as a function of the fraction of galaxies included. The
four lines correspond to splitting up the Ly$\alpha$ survey into $N_{\rm
bin}=2$, 3 and 4 redshift bins (short dashed, long dashed and dot-dashed
respectively). \textit{Right panel:} Comparison between the measurement
error in $\delta T$ at the mean overdensity corresponding to observed
Ly$\alpha$ emitters, with our model for the level of variation of
temperature with overdensity ($C=10$). The errors have been plotted around
the mean expected from the model. The 3 sets of points correspond to
splitting up the Ly$\alpha$ survey into $N_{\rm bin}=2$, 3 and 4 redshift
bins, and the values shown correspond to the minimum value of $\Delta T_{\rm
diff}$ as a function of the fraction of the brightest galaxies used,
$F_{\rm gals}$. The line styles are as per the left panel. The results were
computed assuming experimental parameters that correspond to one SDF,
combined with 1000 hours of integration on the MWA-LFD. }
\label{fig6}
\end{figure*}

\subsection{Ly$\alpha$ emitters as tracers of overdense regions}

Strong clustering of massive sources in overdense regions implies that these sources should trace the higher density regions of IGM. In this section we compute the distribution of overdensities on a scale $R$ that are centered on galaxies of mass $M$. These overdensities are larger than average since galaxies preferentially form in overdense regions. 

The likelihood of observing a galaxy at a random location is proportional
to the number density of galaxies. At small values of large scale
overdensity $\delta$, this density is proportional to $[1+\delta
b(M,z)]$. 
More generally, given a large scale overdensity $\delta$ on a scale $R$,
the likelihood of observing a galaxy may be estimated from the
Sheth-Tormen~(2002) mass function as
\begin{equation}
\label{LH}
\mathcal{L}_{\rm g}(\delta) = \frac{(1+\delta)\nu(1+\nu^{-2p}) e^{-a\nu^2/2}}{\bar{\nu}(1+\bar{\nu}^{-2p})e^{-a\bar{\nu}^2/2}}, 
\end{equation}
where $\nu=(\delta_{\rm c}-\delta)/[\sigma(M)]$ and $\bar{\nu} =
\delta_{\rm c}/[\sigma(M)]$. Here $\sigma(M)$ is the variance of the
density field smoothed with a top-hat window on a mass scale $M$ at
redshift $z$, and $a=0.707$ and $p=0.3$ are constants. Note that here as
elsewhere in this paper, we work with overdensities and variances computed
at the redshift of interest (i.e. not extrapolated to
$z=0$). Equation~(\ref{LH}) is simply the ratio of the number density of
halos in a region of over-density $\delta$ to the number density of halos
in the background universe. This ratio has been used to derive the bias for
small values of $\delta$ (Mo \& White~1996; Sheth, Mo \& Tormen~2001). For
example, in the Press-Schechter~(1974) formalism we write
\begin{eqnarray}
\nonumber \mathcal{L}_{\rm g}(\delta) &=&
(1+\delta)\left[\frac{dn}{dM}(\bar{\nu}) +
\frac{d^2n}{dMd\nu}(\nu)\frac{d\nu}{d\delta}\delta\right]\left[\frac{dn}{dM}(\bar{\nu})\right]^{-1}
\\&\sim&
1+\delta\left(1+\frac{\nu^2-1}{\sigma(M)\nu}\right)\equiv1+\delta b,
\end{eqnarray}
where $(dn/dM)(\bar{\nu})$ and $(dn/dM)(\nu)$ are the average and perturbed mass functions, and $b$ is the bias factor.
Utilising Bayes theorem, we find the a-posteriori probability distribution
for the overdensity $\delta$ on the scale $R$ given the locations defined
by a galaxy population. We obtain
\begin{equation}
\left.\frac{dP}{d\delta}\right|_{\rm gal}\propto \mathcal{L}_{\rm
g}(\delta)\frac{dP_{\rm prior}}{d\delta},
\end{equation}
where $\frac{dP_{\rm prior}}{d\delta}$ is a Gaussian of variance
$\sigma(R)$.

In Figure~\ref{fig5} we show the differential (left panel) and cumulative
(right panel) distributions of overdensity $\delta(R_{\rm beam})$ at the position of a
Ly$\alpha$ emitting galaxy at $z=6.57$. The overdensities refer to a
density distribution that has been smoothed on a scale $R_{\rm beam}$, and
the Ly$\alpha$ emitters were assumed to have masses of $M=10^{10}M_\odot$
(upper row) and $M=10^{11}M_\odot$ (lower row). In each case the 4 lines
correspond to values of $N_{\rm bins}=1$, 2, 3 and 4. The angular scales corresponding to these cases are $\theta_{\rm beam}=8.3'$, $4.2'$, $2.8'$, and $2.1'$. The means of these
distributions are each greater than zero, with the departure greater in the
case of smaller beam-size (which has larger intrinsic variance).

Not every galaxy will be observed in an overdense region. However we are
interested in correlating the 21cm signal with galaxy position. The
quantity of interest is therefore the distribution of the mean overdensity
as measured using samples of $N_{\rm gal}$ galaxies. For $N_{\rm gal}=30$
and $M=10^{10}M_\odot$, we find that the mean overdensities sampled by
galaxies are $\langle\delta\rangle_{\rm gal}=0.037\pm0.015$, $0.10\pm0.02$,
$0.17\pm0.03$, $0.22\pm0.04$ for $N_{\rm bins}=1$, 2, 3 and 4 respectively. 
These values are each significantly in excess of zero. Mass conservation
implies that regions devoid of galaxies must be underdense.

\subsection{The sensitivity to brightness temperature fluctuations.}

Results for $\Delta T_{\rm diff}$ are plotted in the left panel of
Figure~\ref{fig6}, assuming experimental parameters that correspond to one
SDF, combined with 1000 hours of integration using the MWA-LFD. Here three
lines are plotted corresponding to $N_{\rm bin}=2,3$ and 4 (bottom to
top). The case of $N_{\rm bin }=1$ leads to a near absence of empty regions,
and hence a large uncertainty. The Ly$\alpha$ emitters in the SDF cover an order
of magnitude in luminosity. The noise is therefore plotted as a function of
the fraction of the brightest galaxies used ($F_{\rm gals}$). The noise
level can be reduced as the square-root of the integration time, in
proportion to the collecting area or number of LFD units, and as the
square-root of the number of SDF equivalents surveyed.

First let us suppose that the ionization fraction was uniform across the
whole IGM. In this case the brightness temperature of regions surrounding
galaxies would be greater than average by a factor
$(1+\frac{4}{3}\langle\delta\rangle_{\rm gal})$. Similarly, those regions
without Ly$\alpha$ emitters have $\langle \delta\rangle_{\rm no gal}<0$,
and hence brightness temperatures that are lower than average by a factor
$(1+\frac{4}{3}\langle\delta\rangle_{\rm nogal})$. We define the difference
in brightness temperature between regions that contain and do
not contain galaxies to be $\delta T$. The intrinsic (i.e. uniform
ionization fraction) difference in brightness temperature between regions
containing and not containing galaxies would therefore be
\begin{eqnarray}
\label{intrins}
\nonumber \delta T_{\rm int} &=& 22\mbox{mK}x_{\rm
HI}\frac{4}{3}\left(\langle\delta\rangle_{\rm
gal}-\langle\delta\rangle_{\rm nogal}\right) \\ &\approx&
4\mbox{mK}\left(\frac{\langle\delta\rangle_{\rm
gal}-\langle\delta\rangle_{\rm nogal}}{0.3}\right)\left(\frac{x_{\rm
HI}}{0.5}\right),
\end{eqnarray}
where $x_{\rm HI}$ is the average neutral fraction. This offset must be
subtracted from any measured difference to find the difference in
ionization state between over and under-dense
regions. Equation~(\ref{intrins}) represents a value of $\delta T$
corresponding to an exponent of $\beta=0$ in the relation between
overdensity and neutral fraction (see Equation~\ref{neut}).

From Equation~(\ref{temp}) we find the uncertainty ($\Delta\beta$) in the exponent $\beta$ given an uncertainty in the brightness temperature ($\Delta T_{\rm diff}$) of the overdense regions
\begin{equation}
\Delta \beta \sim 0.25 \left(\frac{\Delta T_{\rm diff}}{1\mbox{mK}}\right) \left(\frac{x_{\rm HI}^R}{0.5}\right)^{-1}\left(\frac{\langle\delta\rangle_{\rm gal}-\langle\delta\rangle_{\rm nogal}}{0.4}\right)^{-1}.
\end{equation}
Figure~\ref{fig6} suggests that $\Delta T_{\rm dif}\sim1$mK will be
achievable with first generation instruments, combined with Ly$\alpha$
surveys of comparable size to those already performed. Given a measured
uncertainty of $\Delta T_{\rm diff}\sim1$mK, we can determine the exponent
$\beta$ to $\Delta\beta=\pm0.25$. This implies that we could easily tell the
difference between reionization scenarios that predict $x_{\rm HI}\propto
(1+\delta)^{-3}$ (i.e. the relation similar to that implied by our model), and a scenario where ionization was uniform with $x_{\rm HI}\propto const.$

Finally, as an example, we compare the expected brightness temperature
noise and overdensity estimates (for experimental parameters that
correspond to one SDF, combined with 1000 hours of integration using the
MWA-LFD) to our model calculation of brightness temperature as a function
of large scale overdensity (with $C=10$). Three curves are shown in the right panel of
Figure~\ref{fig6}, with mock data points showing the estimated error
over-plotted for comparison (here the error bars have been computed for the
value of $F_{\rm gals}$ that provides the smallest value of $\Delta T_{\rm
diff}$). The three sets of points and curves correspond to splitting up
of the Ly$\alpha$ survey into $N_{\rm bin}=2$, 3 and 4 redshift bins. The
model curves have been computed assuming a scale for the overdensities that
corresponds to the beam size ($R_{\rm beam}$) in each case. Note that where $N_{\rm bins}$ is smaller ($\sim2$), the beams containing galaxies fill roughly half of the survey volume, so that $\langle\delta\rangle_{\rm gal}\approx -\langle\delta\rangle_{\rm nogal}$. Conversely, where $N_{\rm bins}$ is larger, the beams containing galaxies fill a small fraction of the survey volume. In this case $\langle\delta\rangle_{\rm gal} > -\langle\delta\rangle_{\rm nogal}$. This example shows that even at
this late stage of reionization, first generation surveys will be able to
detect the correlation of galaxies with 21cm emission.  In the future,
larger surveys and instruments will allow substantial improvements. For
example, an array with 10 times the MWA-LFD collecting area, combined with
4 SDF's would reduce the uncertainty in $\beta$ by a factor of $\sim20$.

\section{Summary}

In this paper we have calculated the expected cross-correlation between the
distribution of galaxies and the intergalactic 21cm emission at high
redshifts.  We constructed a simple model for reionization that accounts
for both galaxy bias and an enhanced recombination rate in overdense
regions, and used this model to compute the ionization fraction as a
function of large scale overdensity in the IGM. Our model predicts that
overdense regions will be ionized early as a result of their biased galaxy
formation.  This early phase of reionization in overdense regions leads to
anti-correlations between the 21cm emission and the overdensity of baryons,
and between the 21cm emission and the overdensity of neutral hydrogen. In
addition, because galaxies are biased towards overdense regions, our model
also predicts an anti-correlation between 21cm emission and the galaxy
population.

To explore the detectability of any correlation between 21cm emission and
galaxy properties, we also constructed a simple model for the Ly$\alpha$
emission corresponding to a dark-matter halo of a given mass, and hence for
the Ly$\alpha$ luminosity function. We compared this model to an existing
Ly$\alpha$ survey in the Subaru Deep Field. Through this comparison we
demonstrated that current surveys probe galaxy masses that are larger than
$10^{10}M_\odot$. Due to their biased formation, galaxies of this mass are
highly clustered in overdense regions of the IGM. We have shown that by
comparing 21cm emission from regions near observed galaxies to those away
from observed galaxies, future redshifted 21cm observations will be able to
test the generic prediction that overdense regions are reionized first.

{\bf Acknowledgments} The research was supported by the Australian Research
Council (JSBW) and Harvard University grants (AL).

\newcommand{\noopsort}[1]{}

\label{lastpage}
\end{document}